\documentclass[doublecol]{epl2} 
\usepackage{amsmath}

\title{Quark matter equation of state and stellar properties}
\shorttitle{Quark matter EoS and stellar properties} 

\author{J. R. Torres \and D. P. Menezes 
}
\shortauthor{J. R. Torres and D. P. Menezes}

\institute{                    
  Depto de Física CFM, Universidade Federal de Santa Catarina Florianópolis - CP 476, CEP 88.040-900 Florianópolis SC Brazil\\
}

\abstract{
In this paper we study strange matter by
investigating the stability window within the QMDD
model at zero temperature and check that it can explain the very
massive pulsar recently detected.  We compare our results with the
ones obtained within the MIT bag model and see that the QMDD model
can explain larger masses, due to the stiffening of the equation of
state.
}

\begin{document}

 \maketitle

\section{Introduction}

Neutron star is a very dense object believed to be the remnant of a supernova explosion. Its
estimated mass lies around $\left( 1-2M_{\odot }\right)$, its radius is of the order $10$ $\mbox{km}$
and its temperature of the order of $10^{11}K$ at birth, cooling
rapidly to about $10^{10}K$ by emitting neutrinos. In most
physical models for neutron stars, the star is composed only of hadrons and
leptons, whose stellar matter is a mixture of degenerate neutrons, protons and electrons.
In all stellar models, the structure matter depends on the
assumed equation of state (EOS), which is derived from effective
models. One of the uncertainties related to 
neutron star properties is the ground state of
nuclear matter. In most models, hadrons are assumed to be the true ground
state of the strong interaction. Since the proposal by Witten \cite{witten} that strange matter (SM)
may be the actual ground state of baryon matter at high densities, many investigations have been conducted to verify the
veracity of this hypothesis and the implications in several areas of
physics, astrophysics and cosmology with possible consequences in the QCD phase diagram.

SM was first considered in calculations obtained within the MIT bag
model framework \cite{mit}. More sophisticated treatments for SM, based on
the Nanbu-Jona-Lasinio \cite{meu} and the color flavor locked phase models
\cite{meu,sergio} also exist in the literature.

In \cite{fowler}, a confinement mechanism 
was introduced by assuming that the quark masses are density
dependent. This model, named QMDD, was then applied to describe SM 
and the related quark star properties in \cite{chakrabarty91}. In the
very same paper, the authors pointed out that the results obtained
with the QMDD model were quite different from the ones obtained 
with the MIT model. Subsequently, in \cite{lugones95}, the authors 
claimed that this difference was due to an incorrect thermodynamical 
treatment of the problem and recalculated the equation of state
showing that an extra term is present in the energy density and
pressure of the system. This extra term results from the dependence 
of the quark masses on the baryonic density. 

An important ingredient in the SM hypothesis is the stability window,
identified with the constant values of the model that are consistent
with the fact that two-flavor quark matter must be unstable (i.e.,
its energy per baryon has to be larger than 930 MeV, which is the
iron binding energy) and SM (three-flavor quark matter) must be
stable, i.e., its energy per baryon must be lower than 930 MeV. 
It was also shown in \cite{lugones95} that the zero pressure density does 
not correspond to the minimum of
the energy per baryon, as is normally the case, because of the quark
mass density dependence. In \cite{wang00}, the author has shown that
the pressure at the density corresponding to the minimum of the energy
density can be zero if it is calculated in a self-consistent way along
the energy density. That calculation requires another extra term in
the grand canonical thermodynamical potential, which mimics 
the MIT bag constant, but
instead of being constant, it increases with the increase of the density.

Still another calculation of the equation of state based on
thermodynamics and also on the general ensemble theory was obtained in
\cite{peng00}. The authors claim that the extra term reported in
\cite{lugones95} is correct in the expression of the
pressure, but should not be present in the energy density equation.
Similar kinds of density dependence are found in  hadronic models
\cite{tw,br} and used to describe protoneutron star
properties in \cite{nossot,nossobr}. Specifically, in \cite{br,nossobr}, the
hadronic masses are density dependent while in \cite{tw,nossot} the
coupling constants of the model are density dependent.
In both approaches, the pressure also carries
a rearrangement term that appears due to the density dependence,
while such a term cancels out in the energy density expressions.
In \cite{peng00}, due to quark confinement and asymptotic freedom,
 another prescription for the quark masses is used and the values for
 the quark current masses are somewhat different from the standart
 ones. Normally $m_{0u}=m_{0d}=5$ MeV and $m_{0s}$ is of the order of
 150 MeV. In \cite{peng00}, $m_{0u}=5$ MeV, $m_{0d}=10$ MeV and
 $m_{0s}=80$ or 90 MeV. Depending on the parametrization used, SM is 
completely stable or becomes metastable.

More recently, the QMDD model was again revisited \cite{su2008}. As already
explained in this Introduction, the grand canonical
thermodynamical potential is not only a function of $T,V$ and
the chemical potential $\mu$, but also depends explicitly on the quark
(or related baryon) density through its mass and this fact caused all
the different treatments described below in the previous papers. As
pointed out in \cite{su2008}, the extra terms for the different
treatments contradict each other and tackling thermodynamics
self-consistently in the QMDD model seems to be a real problem.
Another recipe is then proposed, considering an ideal quasi-particle
system, where the mass can be both density and temperature dependent,
but is taken as constant at fixed values of $T_0$ and $\rho_0$. In
this case, the standard ideal gas expressions are recovered and the
extra terms do not appear. Moreover, the pressure of the system is
always positive and this poses a problem related to the stability of
SM. In order to circunvent this difficulty, the authors have to consider the
physical vaccum.

In 2010, a measurement of the Shapiro delay in the radio pulsar PSR
J1614-2230 yielded a mass of $1.97 \pm 0.04 M_\odot$, the highest mass
of a compact object ever measured.  It is well known that the stellar mass
that can be supported against gravitational collapse depends on the
equation of state used to describe it.  According to \cite{Duarte},
very massive stars, with masses $2 M_\odot \le M \le 2.73 M_\odot$
can be interpreted as compact stars composed entirely of deconfined
quarks and would be quark stars.

The aim of the present work is to check whether this high mass pulsar 
can be described by the QMDD model. A necessary investigation that 
preceeds the use of the equation of state as input to the TOV
equations \cite{tov} is the detailed consideration of the stability window. 
In the present paper we consider quarks with equal chemical 
potentials, closer to the situation in $^{56}Fe$ where matter is almost 
symmetric, than to the situation of charge neutraliy usually analysed, where
electrons, not present in nuclei, are included.
Whenever possible, our
results are compared with the MIT model. In face of all the
discussions on the thermodynamical consistency of different versions
of the QMDD model, we restrict ourselves to the version obtained in
\cite{lugones95}.

\section{Quark matter}

In this section we summarize the models we use to describe the properties of
quark matter. 
The QMDD model is based on a phenomenological approach \cite{lugones95}  
where the dynamical masses of
the three lightest quarks scale inversely with the baryon number density 
\cite{chakrabarty91}, what mimics an interaction among the quarks:

\begin{equation}
m_{u,\overline{u}}^{\ast }=m_{d,\overline{d}}^{\ast }=\frac{C}{3n_{B}},~~~~ m_{s,\overline{s}}^{\ast }=m_{0s,\overline{s}}+\frac{C}{3n_{B}} \mbox{,}
\label{ansatz_massa}
\end{equation}

\noindent
where $C$ is an energy density constant in the zero quark density limit,

\begin{equation}
n_{B}=\frac{1}{3}\left( \rho _{u} +\rho _{d} + \rho _{s}\right) \mbox{,}
\end{equation}

\noindent
is the baryon number density and $m_{0s,\overline{s}}$ is the current mass the of
the $s$ quark. The energy density, the
pressure and the quark density are respectively given by

\begin{equation}
p=\sum_{q}\frac{\gamma _{q}}{2\pi ^{2}}\int_{0}^{\infty }%
\frac{p^2}{E_{q}^{\ast }\left( p\right) }\left[\frac{p^2}{3}-\frac{C}{3n_{B}}\right] \left( f_{q+}^{\ast
}+f_{q-}^{\ast }\right)dp,
\label{pressao_qmdd}
\end{equation}

\begin{equation}
\varepsilon =\sum_{q}\frac{\gamma _{q}}{2\pi ^{2}}\int_{0}^{\infty }%
\frac{p^2}{E_{q}^{\ast }\left( p\right) }\left[E_{q}^{\ast 2}\left( p\right)+\frac{C}{3n_{B}}\right] \left( f_{q+}^{\ast
}+f_{q-}^{\ast }\right) dp,
\label{densidade_energia_qmdd}
\end{equation}

\begin{equation}
\rho _{q}=\frac{\gamma _{q}}{2\pi ^{2}}\int_{0}^{\infty }p^{2}\left(
f_{q+}^{\ast }-f_{q-}^{\ast }\right) dp,
\label{densidade_quarks_qmdd}
\end{equation}

\noindent
where $q=u,d,s,$ and
$m_{q}^{\ast }$ is the effective quark mass. The distribution functions for quarks and
anti-quarks are the Fermi-Dirac distributions
$ f_{q\mp }=1/\left( 1+\exp \left[ \left( E_{q}^{\ast }\left( p\right) \mp \mu
_{q}\right) /T\right] \right)$,
with the chemical potential for quarks (upper sign) and anti-quarks (lower
sign) and $E_{q}^{\ast }\left( p\right) =\sqrt{p^{2}+m_{q}^{\ast 2}}$.
For $T=0$, there are no antiparticles, the chemical potential is equal to the Fermi energy, and
the distribution function for the particles is the usual step function: 
$ f_{q\mp }=\Theta _{q}\left( p_{f}^{2}-p^{2}\right)$ .

In the description of compact stars, both charge neutrality and $\beta
$ equilibrium conditions have to be imposed  \cite{meu}:

\begin{equation}
2\rho _{u}=\rho _{d}+\rho _{s}+3\left( \rho _{e}+\rho _{\mu }\right) 
\label{cn}
\end{equation}
and
\begin{equation}
\mu _{s}=\mu _{d}=\mu _{u}+\mu _{e}\mbox{, \ \ }\mu _{e}=\mu _{\mu }\mbox{.}
\label{beta}
\end{equation}

For the electron and muon pressure, energy density and densities we just replace $q\rightarrow l$ in equations
(\ref{pressao_qmdd}),(\ref{densidade_energia_qmdd}) and (\ref{densidade_quarks_qmdd}), where $l=e,\mu $ and $\gamma_{l}=2$.

We next summarize the main formulae at $T=0$ for both models used in
this paper, the QMDD model \cite{lugones95} and MIT bag model \cite{mit}.

The energy density, the pressure, the quark density and the baryonic
density given in eqs. (\ref{pressao_qmdd}),(\ref{densidade_energia_qmdd})
and (\ref{densidade_quarks_qmdd}) can be rewritten respectively as: 

\begin{equation}
 p=\sum_{q}\frac{\gamma _{q}m_{q}^{\ast 4}}{48\pi ^{2}}F\left(
x_{q}^{\ast } \right)
-B\left( C,m_{q}^{\ast }\right),
\label{eos_param_qmdd_pressao}
\end{equation}

\begin{equation}
\varepsilon =\sum_{q}\frac{\gamma _{q}m_{q}^{ \ast 4}}{48\pi ^{2}}3H\left(
x_{q}^{\ast}
\right) +B\left( C,m_{q}^{\ast }\right),
\label{eos_param_qmdd_densidade_energia}
\end{equation}

\begin{equation}
\rho _{q}=\frac{\gamma _{q}}{6\pi ^{2}}m_{q}^{\ast 3}x_{q}^{3},
\label{eos_param_qmdd_densidade_quarks}
\end{equation}

\begin{equation}
n_{B}=\frac{1}{3}\sum_{q}\frac{\gamma _{q}}{6\pi ^{2}}m_{q}^{\ast
3}x_{q}^{3}\mbox{,}
\label{eos_param_qmdd_densidade_barionica}
\end{equation}

\noindent
where the dynamic term of confinement is given by:

\begin{equation}
B\left( C,m_{q}^{\ast}\right) =\sum_{q}\frac{\gamma _{q}m_{q}^{\ast 4}}{48\pi
^{2}}\frac{C}{n_{B}}\left( \frac{4}{m_{q}^{\ast }}\right) G\left(
x_{q}^{\ast }\right) \mbox{,}
\label{termo de confinamento}
\end{equation}
and the functions

\begin{align}
F\left( x_{q}^{\ast }\right) =&x_{q}^{\ast }\left( x_{q}^{\ast 2}+1\right)
^{1/2}\left( 2x_{q}^{\ast 2}-3\right)\nonumber \\
&+3\ln \left[ \left( x_{q}^{\ast
2}+1\right) ^{1/2}+x_{q}^{\ast }\right] \mbox{,}
\label{funcao_F(X)}
\end{align}

\begin{align}
H\left( x_{q}^{\ast }\right) =&x_{q}^{\ast }\left( x_{q}^{\ast 2}+1\right)
^{1/2}\left( 2x_{q}^{\ast 2}+1\right) \nonumber \\ 
&-\ln \left[ \left( x_{q}^{\ast
2}+1\right) ^{1/2}+x_{q}^{\ast }\right] \mbox{,} 
\label{funcao_H(X)}
\end{align}

\begin{equation}
G\left( x_{q}^{\ast }\right) =x_{q}^{\ast }\left( x_{q}^{\ast 2}+1\right)
^{1/2}-\ln \left[ \left( x_{q}^{\ast 2}+1\right) ^{1/2}+x_{q}^{\ast }\right] \mbox{,}
\label{funcao_G(X)} 
\end{equation}

\noindent
and 

\begin{equation}
x_{q}^{\ast }=\left[ \left( \frac{\mu_q}{m_{q}^{\ast }}\right) ^{2}-1\right]
^{1/2}.
\label{definicao_parametro_x}
\end{equation}

Note that the term $B\left( C,m_{q}^{\ast }\right) $ arises from the
derivatives of the 
grand canonical thermodynamical potential of a Fermi
gas with respect to $m^{*}$ \cite{Duarte}.
This density dependent term leads to a thermodynamical inconsistency, because
the fundamental relation of thermodynamics
$\Omega =-pV$ is violated.
Although the model has this problem, we note that
this density dependent term leads to an empirical model incorporation of the
quark confinement at the low density regime and to asymptotic freedom at very
high densities, as already emphasized in the Introduction.

For the usual MIT bag model \cite{mit} the quark masses are fixed 
as $m_u=m_d=5$ MeV, $m_s=150$ MeV and
its expressions are 
(\ref{eos_param_qmdd_pressao}),(\ref{eos_param_qmdd_densidade_energia}),
(\ref{eos_param_qmdd_densidade_quarks}) where $m^{\ast }$ is simply replaced by $m$ and $B\left( C,m_{q}^{\ast }\right)$ by a bag constant $B$.

\section{Results and conclusions}
\begin{figure*}[ptbh] 
\centering
\includegraphics[width=3.3in]{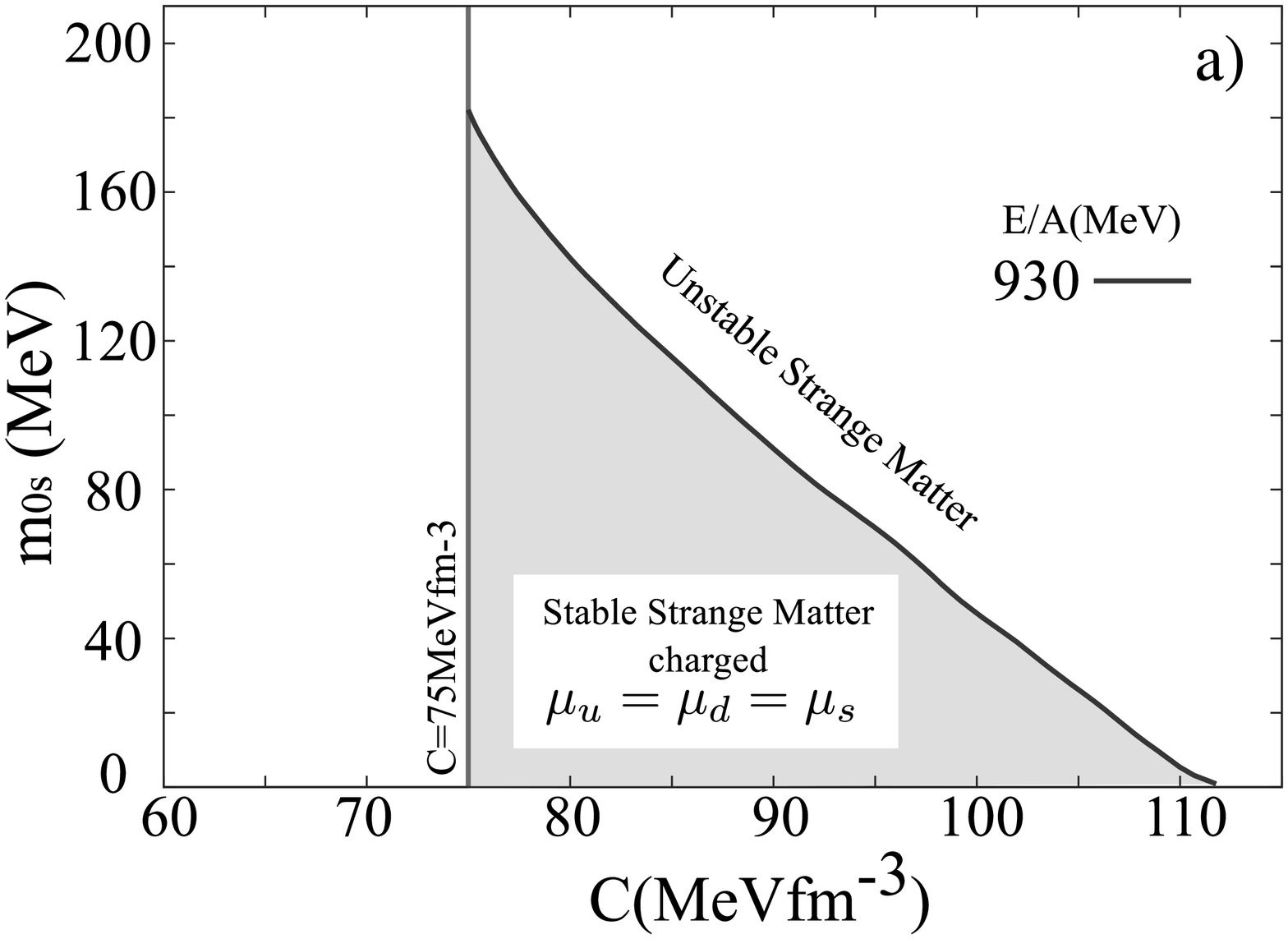}
\includegraphics[width=3.3in]{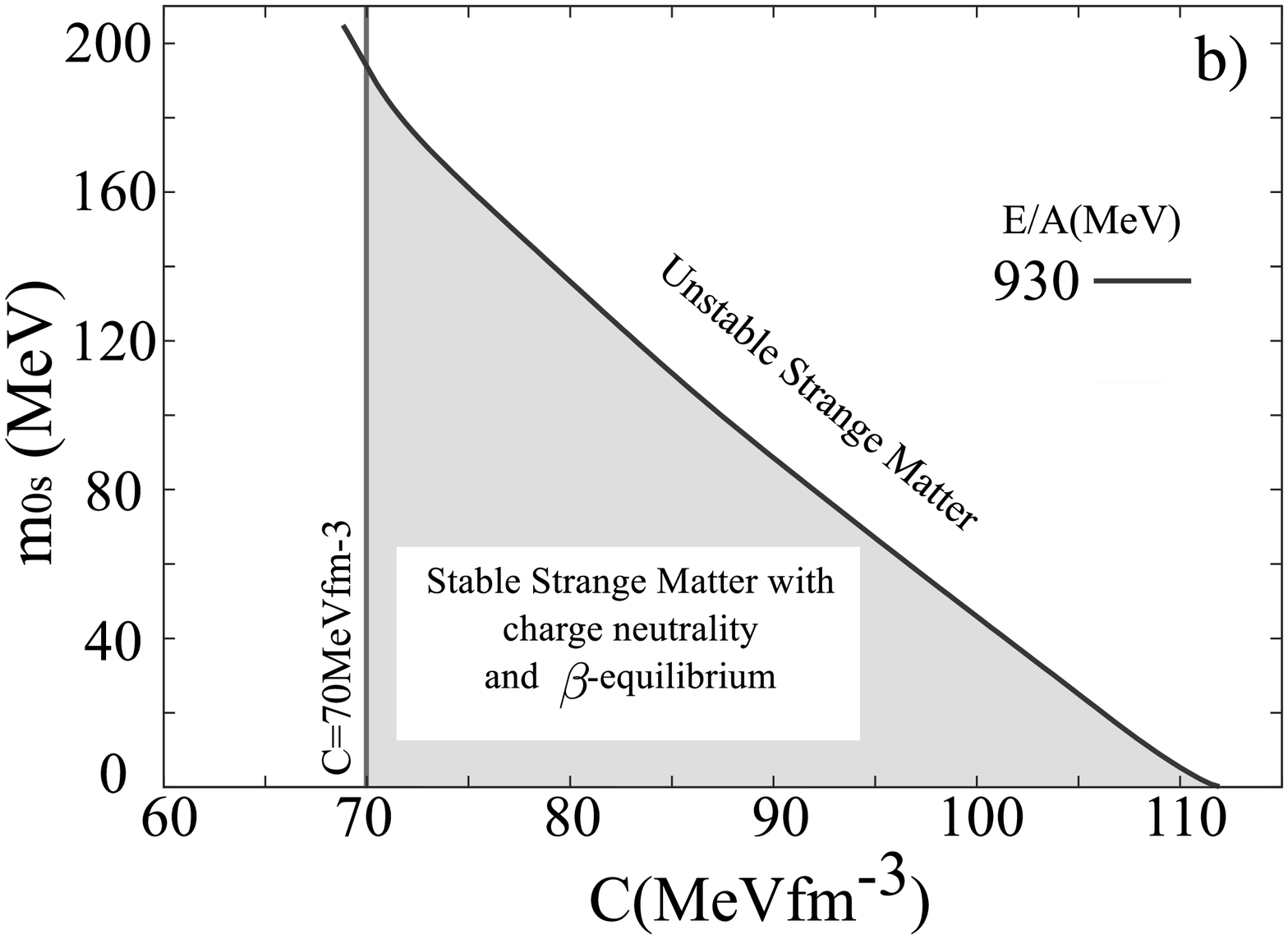}
\caption{SM stability window obtained with the QMDD model at $T=0$. 
The points in the figures show the maximum values ​​for $C$ when $m_{0s} = 150$MeVis fixed.}
\label{f.1}
\end{figure*}

\begin{figure*}[ptbh] 
\centering
\includegraphics[width=3.3in]{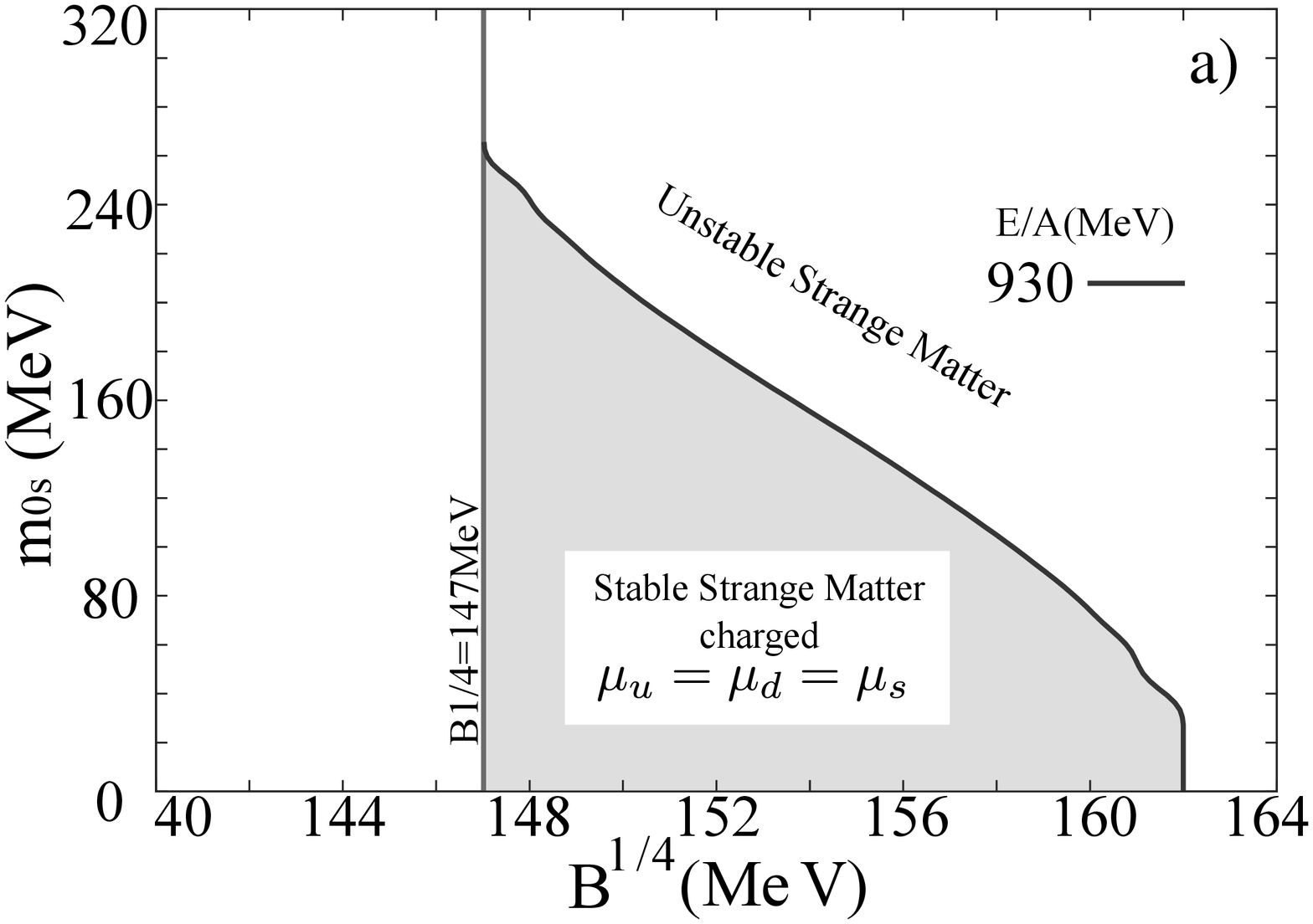}
\includegraphics[width=3.3in]{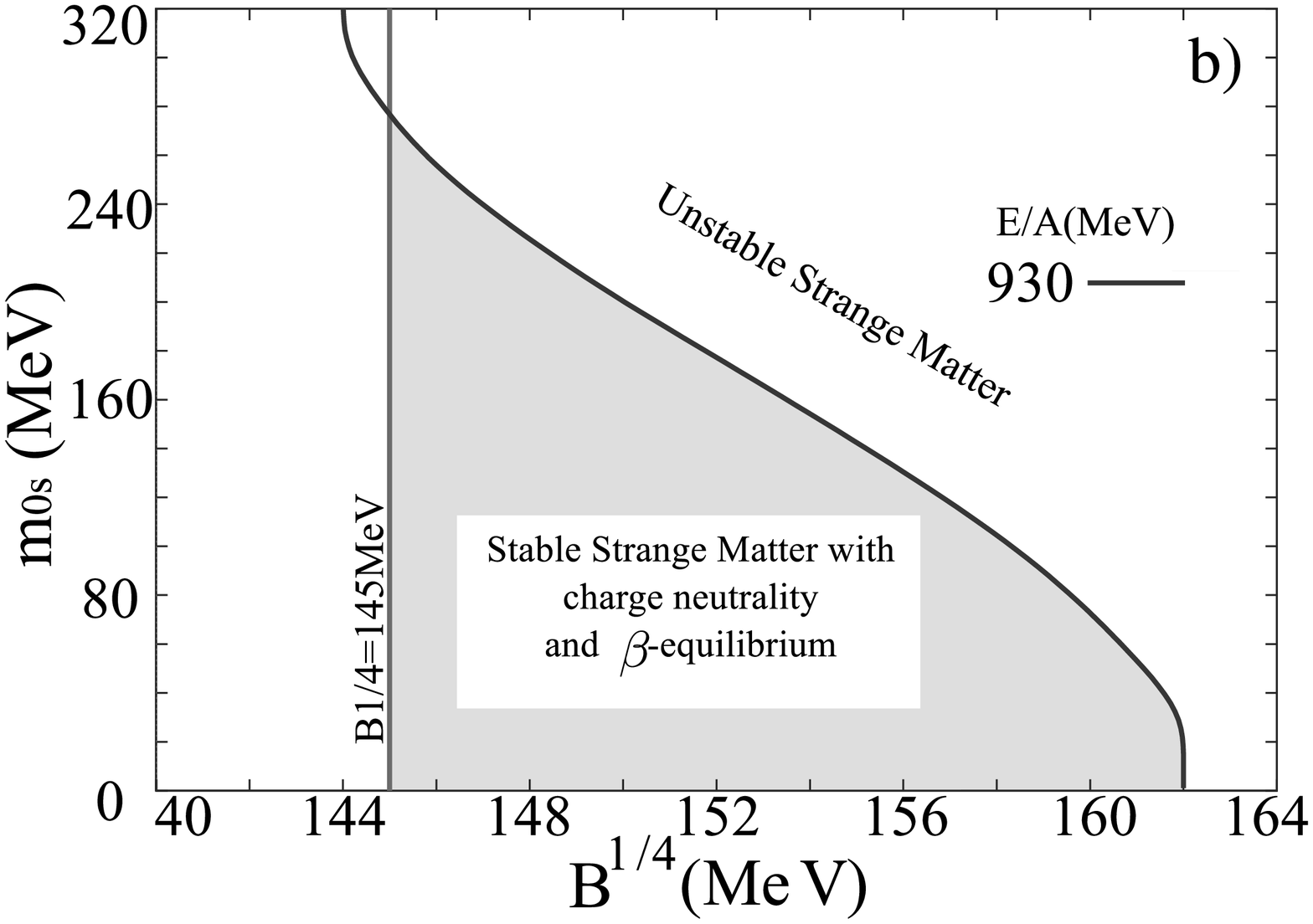}
\caption{SM stability window obtained with the MIT model at $T=0$. 
The points in the figures show the maximum values ​​for B$^{1/4}$
when $m_{0s} = 150$MeVis fixed.}
\label{f.2}
\end{figure*}

\indent
We now want to establish the conditions under which SM is the true ground state \cite{witten,lugones95}.
In principle, the theory of the strong interaction should contain the answer to the question of whether strange matter is stable. 
Unfortunately, as we all know, QCD is still far from being completely solved. We need to rely on effective models to test the idea of SM.
Hence, we next study strange matter via two different effective models and find a sizeable region of parameter 
space for which strange matter is stable. For the QMDD case, these parameters are $C$ and the current strange quark mass $m_{0s}$
(see equation (\ref{ansatz_massa})). For the MIT bag model, these
parameters are the well-known $B^{1/4}$ and and current strange quark mass $m_{0s}$.
Strange matter is stable at zero pressure if its energy density per 
baryon density is lower than the energy per baryon 
of $^{56}\mbox{Fe}$,
 $\left(\frac{\varepsilon}{n_{B}} = \frac{E}{A}\right)_{SM}\leq930\mbox{MeV}$ 
while the two-flavor quark matter must be unstable (i.e.,
its energy per baryon has to be larger than 930 MeV).
Notice that, if finite size effects
are taken into account, this value is around 4 MeV lower, i.e. 934 MeV 
\cite{mit}, 
but the final stability window remains practically unchanged. The
numbers are discussed below.
The energy per nucleon has been calculated numerically for 2QM and SM 
respectively. Therefore, 
SM is stable in the shaded region shown in Fig.~\ref{f.1},
for the QMDD model and in Fig.~\ref{f.2} for the MIT model.   
The lower limit, vertical straight line, is due to the requirement that two flavor quark matter is not absolutely stable.
In both figures one can see that we have allowed the strange quark
mass to vary considerably. We have checked that the results are slightly 
different if we consider
matter with identical quark chemical potentials, corresponding to
symmetric matter in the 2QM or charge neutral matter in 
$\beta$-equilibrium, as
expected in stars. For both models, the stable region is larger if  
the conditions given in eqs.(\ref{cn}) and (\ref{beta})
are imposed. This
is the situation which is usually analysed in the literature, but one
has to bear in mind that stable nuclear matter (as in iron) 
does not obey these conditions and no electrons are present.
Actually, its proton fraction is $Y_p=0.46$, very
close to symmetric matter and this is the reason why we have chosen to
analyse matter with equal quark chemical potentials.
If we consider finite size effects and take $m_{0s}=150$
  MeV, the border of the stability line moves from $C=80.1$ to 79.2 MeV/fm$^3$ in the
  QMDD model and from $B^{1/4}=155.1$ to 155.8 in the MIT model.

\begin{figure*}[ptbh] 
\centering
\includegraphics[width=3.0in]{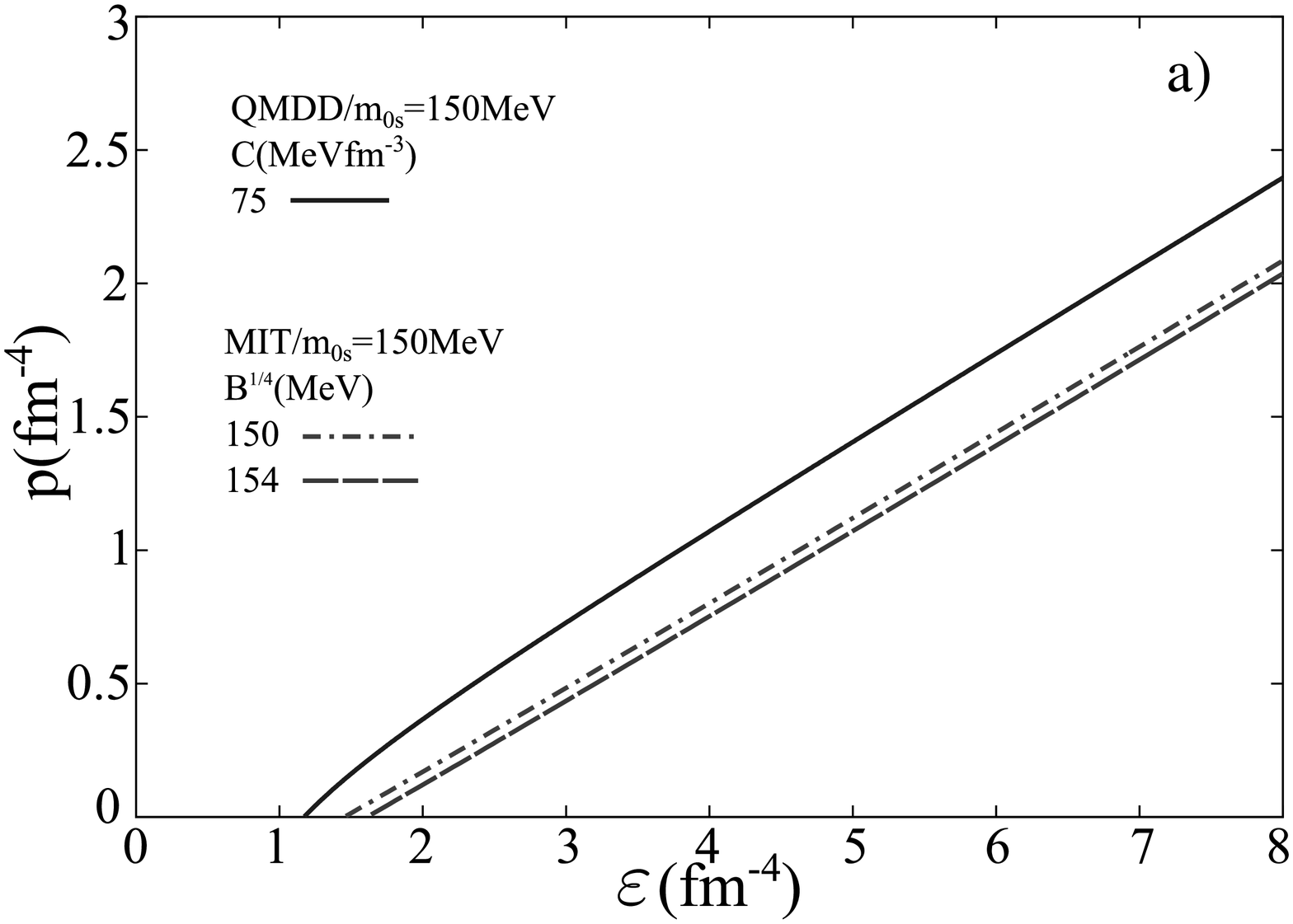}
\includegraphics[width=3.0in]{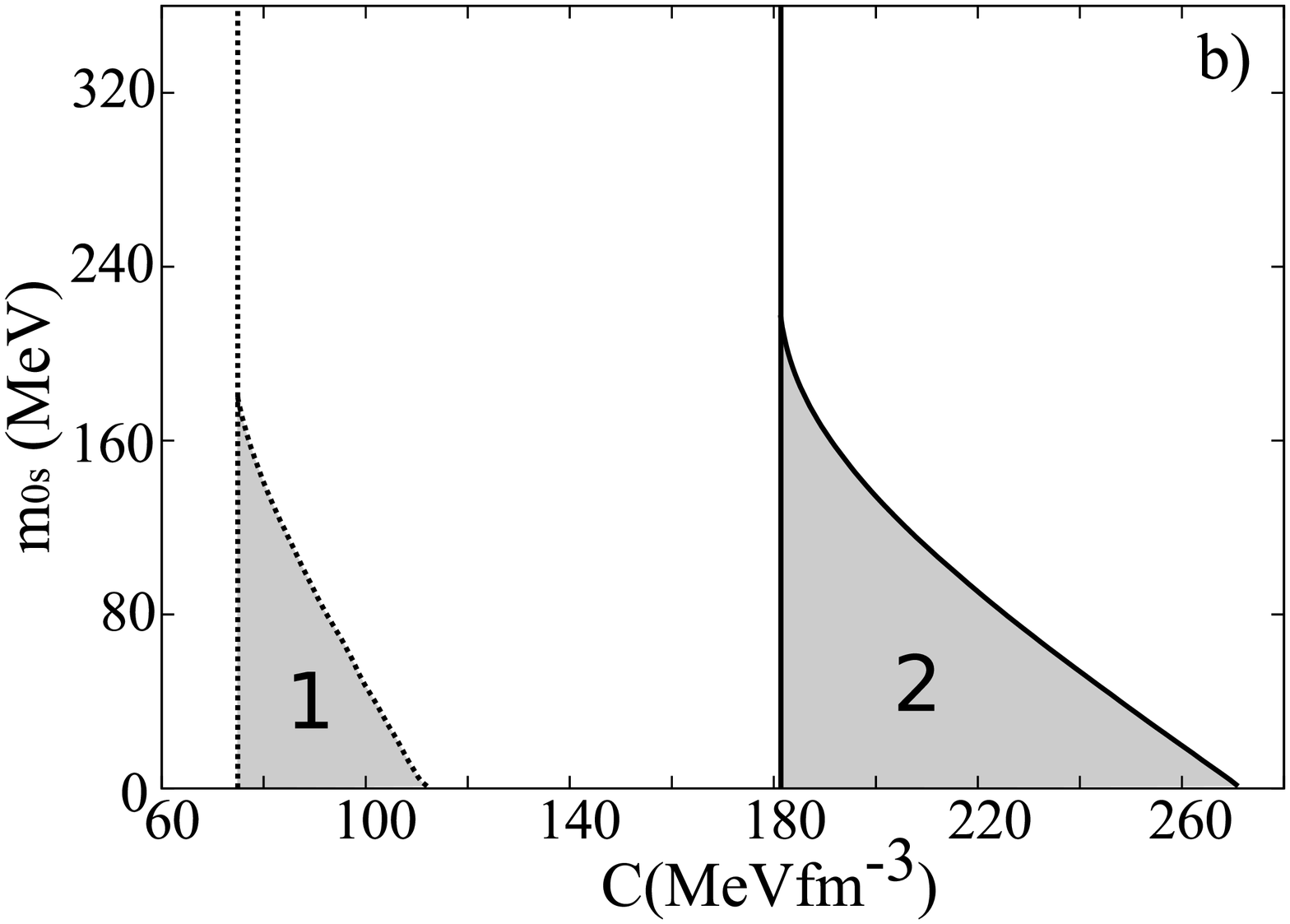}
\caption{In a) we show strange matter equations of state for 3 parameter choices
  and in  b) we compare the
    stability window for two different versions of the QMDD model.}
\label{f.3}
\end{figure*}

In Fig.~\ref{f.3}a, we plot the equations of state (EoS) obtained for two
constants in the MIT model and one situation in the QMDD model, chosen
from the values which satisfy the SM stability condition.
For these choices of parameters, the QMDD EoS is reasonably stiffer.
The MIT bag model presents a linear dependence of the pressure against
the energy density as shown in Fig.~\ref{f.3}. The parameter $B$
is just a linear parameter in the EOS for the MIT case. 
This is not true for the QMDD model, 
because the pressure as a function of the energy density is more complicated, as seen in Fig.~\ref{f.3}.
This non-trivial behavior is due to $B\left( C,m_{q}^{\ast }\right)$ in equation (\ref{termo de confinamento}).
If we consider the thermodynamic treatment used in
  \cite{peng00}, where the zero pressure point corresponds to the
  minimum of the energy per baryon, the stability window modes to larger parameters, as
  seen in Fig.  \ref{f.3}b and  the equations of state become softer.

As stated in the Introduction, in order to study hydrostatic equilibrium of 
a family of quark stars
we have to solve numerically the Tholman-Oppenheimer-Volkoff equations
(TOV) \cite{tov}. The mass radius relation is obtained from the solution 
of these coupled differential equations. 

\begin{figure*}[ptbh] 
\centering
\includegraphics[width=4.in]{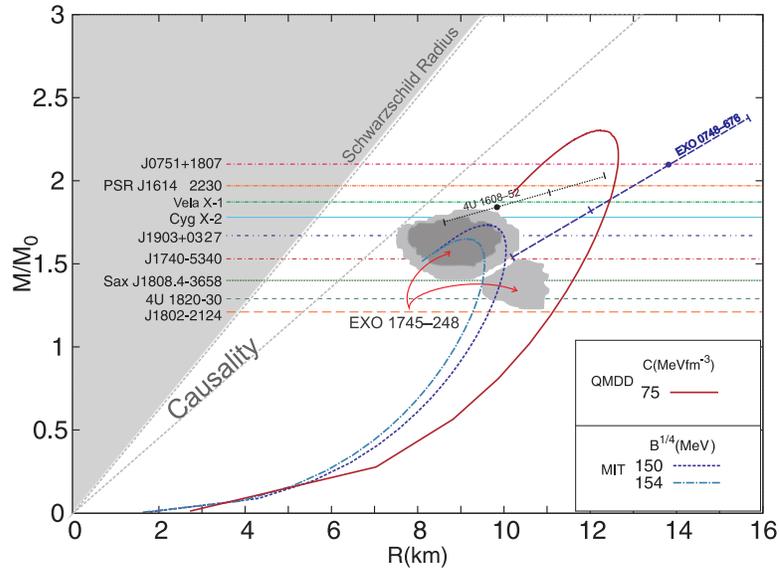}
\caption{Mass-radius relations for different strange stars.}
\label{f.4}
\end{figure*}

\begin{figure*}[ptbh] 
\centering
\includegraphics[width=3.5in]{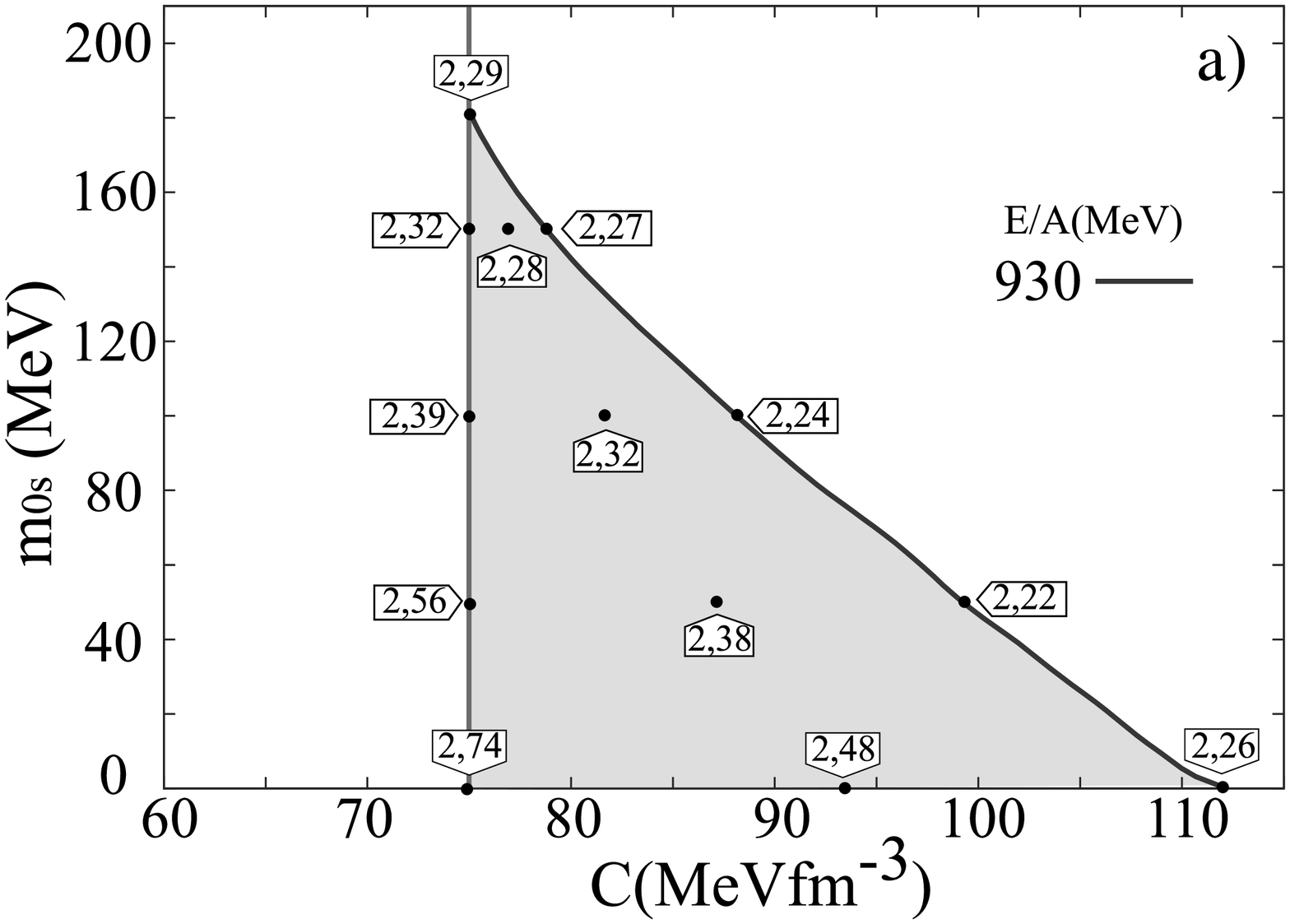}
\includegraphics[width=3.5in]{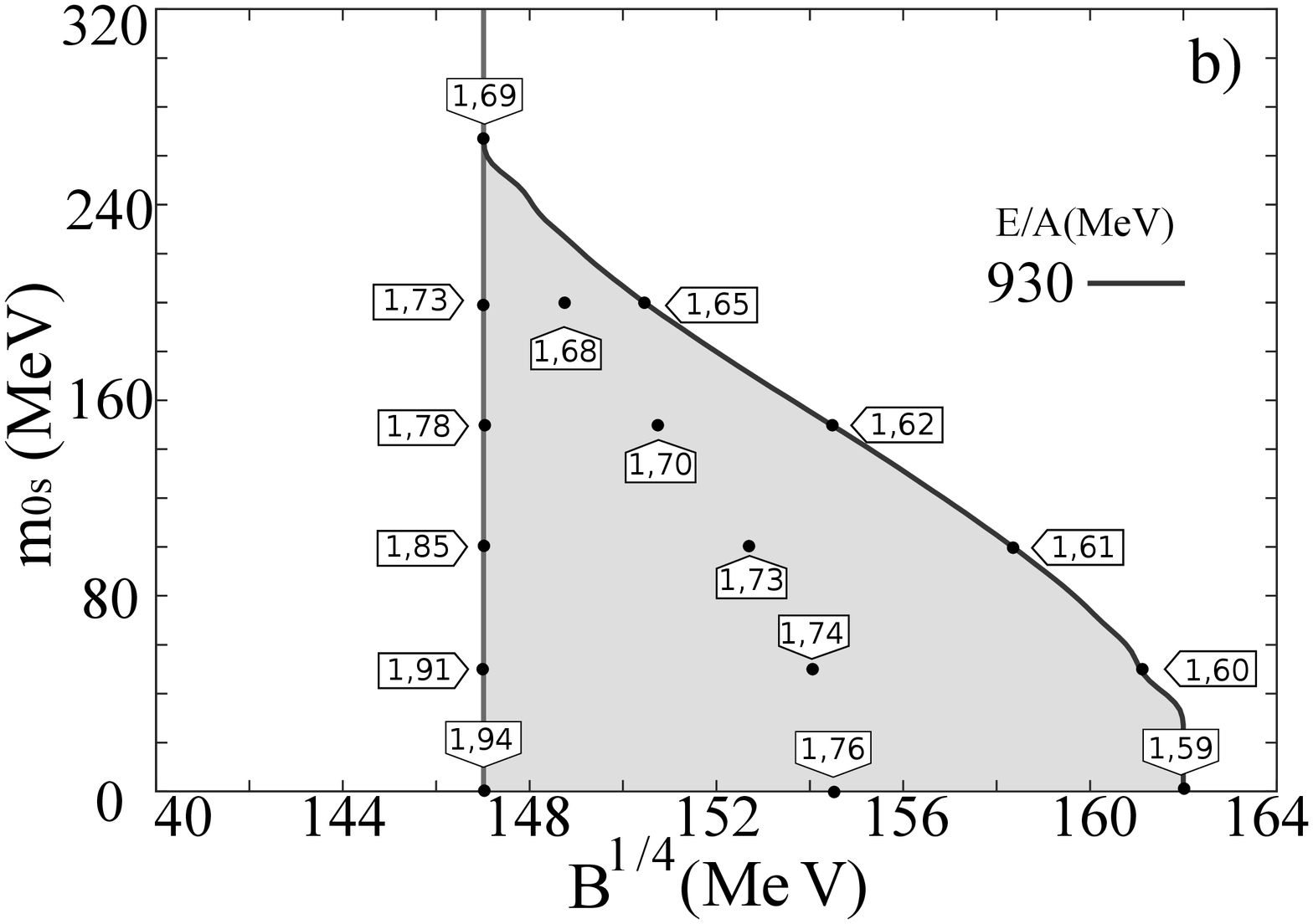}
\caption{Inside the flags we show the maximum masses
    $(\mbox{M/M}_{\odot})$ for the corresponding parameter sets
    inside the stability window for 
a) the QMDD and b) the MIT models respectively.}
\label{f.5}
\end{figure*}

The determination of small radii as the ones expected in pulsars
from observations based on luminosity and temperature is full of
uncertainties \cite{sulei}. Hence, although X-ray astronomy has provided
some simultaneous determination of masses and radii from X-ray
bursters \cite{Ozel}, they have to be taken with care. 
The integration of the TOV equations gives quark stars
whose mass-radius relation is shown in Fig.~\ref{f.4}. 
In this figure the horizontal lines are the masses of some well-known
pulsars extracted from \cite{Zhang, Demorest}. The lower and upper limits of the
masses and radii of EXO 0748-676 and 4U 1608-52 are also displayed. 
The shaded clouds refer to the $1 \sigma$ and $2 \sigma$ confidence
ellipse of the results obatined in \cite{Ozel} for the EXO 1745-248.
We notice that the  QMDD model can certainly describe 
compact objects which are more massive than  the MIT model, as already
expected from its harder EoS.  Nevertheless, it fails to describe
pulsars with low radii, which can be described by the MIT model. One, however,
has to bear in mind the uncertainties discussed in \cite{sulei}.
In Fig. \ref{f.5}, we show the maximum masses obtained in
  different points of the stability window for both models. They
  increase with the decrease of the  strange quark mass and change
  vary little for the paramenters on the right border line. For the
  present EOS, the QMDD model yields sistematically larger maximum
  masses than the MIT model.

All the equations of state obtained for the 
QMDD model discussed in the Introduction rely on derivatives of the 
grandcanonical thermodynamical potential, 
which are truncated at different arbitrary points. To obtain the correct 
expressions, an alternative calculation can be done via the energy-momentum 
tensor as \cite{tw,nossot,nossobr}. This work is in progress.

In a forthcoming paper, we will analyse carefully the stability
windown for protoquark stars described by quark matter at finite
temperature (or fixed entropy). 
The choice of appropriate parameters in the search for stable SM 
at finite temperature systems requires, in addition to the 
investigation presented in this paper, a carefull study of the free energy
per baryon (${\cal F} = \varepsilon - TS$, where $S$
is the entropy density of the system), in contrast to previous works 
\cite{chmaj89, chakrabarty93, lugones95t,su2002}, where the 
energy density per baryon was used.
Of course, the lower limits, given by the straight lines on the left 
of figures \ref{f.1} and \ref{f.2} remain the same because the 
$^{56}Fe$ ground state is a zero temperature system, where the binding 
energy and the free energy per baryon coincide.

\acknowledgments 

This work was partially supported by CNPq and CAPES. The authors acknowledge 
fruitful discussions with Prof. S\'ergio Barbosa Duarte.

\end{document}